# Driven charge density modulation by spin density wave and their coexistence interplay in SmFeAsO: A first-principles study

Toktam Morshedloo[1], Ali Kazempour[1,2*], Hamideh Shakeripour[3], S. Javad Hashemifar[3], Mojtaba Alaei[3]

[1] *Department of Physics, Payame Noor University, PO BOX 119395-3697, Tehran, Iran*
[2] *Nano Structured Coatings Institute of Yazd Payame Noor University, PO Code: 89431-74559, Yazd, Iran*
[3] *Department of Physics, Isfahan University of Technology, Isfahan, 84156-83111, Iran*

## Abstract

We use density functional theory to investigate effects of spin-orbit coupling and single-stripe-type antiferromagnetic (sAFM) ordering on the crystal structure and electronic properties of SmFeAsO. The results indicate that AFM ordering causes the crystal structure transition from tetragonal to orthorhombic, along with increase in the height of As atoms from Fe layer due to magnetostriction. It also leads to the opening of partial band gaps, the emergence of a prominent peak near Fermi energy ($E_F$) in the density of states (DOS) and a reduction in Fermi surface nesting. The study finds a correlation between the calculated area under the DOS curve, spanning from $E_F$ to the first peak above it, and the optimal electron-doped concentration required for inducing superconductivity in SmFeAsO. The findings suggest that spin and charge density waves can play important roles in the superconducting mechanism of Fe-based superconductors. Our calculations demonstrate that spin-orbit coupling reduces electronic correlation.

---

[*] kazempour@pnu.ac.ir,
Tel: +98-03535216133

# INTRODUCTION

Since the iron pnictide layered superconductors were discovered, a hope was raised that these materials would finally reveal the mechanism of the superconductivity in the high transition temperature ($T_C$) superconductors (SC) [1, 2, 3, 4]. Numerous empirical and semi-empirical theories have been developed to understand the origin of the pairing mechanism [1, 3, 5, 6]. It is desirable that a fully parameter-free theory be developed which is able to reproduce the essential properties of the SC (including the critical temperature, complex gap function and excitation spectrum) based on the only knowledge of the atomic constituent and chemical structure [1]. Density functional theory (DFT) is the method for ab-initio calculations of electronic, magnetic and structural properties in the normal state. Moreover, DFT has been generalized to describe the superconducting phase as superconducting density functional theory (SCDFT) [4, 7, 8, 9]. It was observed that a satisfactory description of the normal state (metal and undoped parent compound of SC) affects the predictive power of the theoretical approach [3, 4]. In other words, the SCDFT scheme is based on the assumption of a second order phase transition between the SC phase and the normal metallic parent compound [3]. Therefore, $T_C$ and other essential properties of the SC can be estimated through the electronic structure of the metallic non-superconducting phase as a starting point to act-on with a pairing field computed from first principle [3]. Hence, accurately calculating the electronic and magnetic properties of metallic non-superconducting phase is the basic starting point of the SCDFT that we do for SmFeAsO as a member of pnictide superconductors in the present work. Experimental facts indicate that only the 1111-type Fe-As-based superconductors exhibit a high-$T_C$ above 50 K which belongs to SmFeAsO$_{1-x}$F$_x$ [4, 10, 11, 12]. The parent compounds of the FeAs-1111 family, including SmFeAsO, have a quasi-two-dimensional electronic structure which is distinguished by the nested electron and hole Fermi surface (FS). The magnetic ordering is considered to be one of the consequences of nesting on the Fermi surface [13, 14]. Based on the SCDFT calculations for Fe-based superconductors, a spin fluctuation model resulting from Fermi surface nesting between hole and electron pockets was proposed as a superconducting mechanism [3]. Moreover, it was observed a charge density modulation coexisting with the superconducting phase in Fe-based superconductors [15].

Accordingly, it is wise to investigate the interplay between spin-orbit coupling (SOC) effect, crystal structure and electronic and magnetic properties of SmFeAsO. Although, a lot of attempts have been performed to study 1111-type Fe-based superconductors experimentally as well as theoretically [3, 4, 5, 11, 14, 16, 17], but no detailed ab-initio investigations have been carried out to survey the effect of SOC and magnetic ordering on electronic correlation and crystal structure of SmFeAsO.

The SmFeAsO compound undergoes a phase transition from a high-temperature tetragonal (P4/nmm) to a low-temperature orthorhombic (Cmma) structure as the temperature decreases. Furthermore, an orthorhombic-tetragonal structural transition occurs during the transition from a normal-state to a superconducting state by doping [18, 19, 20]. On the other hand, the charge density wave (CDW) formation depends on lattice distortion as an essential element [21]. Therefore, it is crucial to study both the tetragonal (Tet) and orthorhombic (ORT) crystal structure simultaneously. It provides opportunity to evaluate solely effects of magnetic ordering and SOC interaction on the electronic properties of SmFeAsO with considering the distortion from Tet to ORT structure [22].

Since, SmFeAsO compound with transitional Fe and lanthanide Sm ions, is a correlated system, we take the coulomb repulsion into account via applied Hubbard interaction (U) in our calculations through full potential method. In the following, we study the effect of SOC on the electronic correlation through calculating the value of Hubbard U in the presence of SOC interaction. The paper has the following outline: After the computational details in the next section (Sec. 1), we present results and discussions in Sec 2. At first, the calculated characters of the crystal structure are exhibited and inspected in the presence and absence of the magnetic ordering as well as with and without considering SOC (Sec. 2. 1.). Then, we analyze the electronic properties in the various conditions mentioned former through the investigation of partial and total density of states (PDOS and DOS), electronic band structure and Fermi surface (Sec. 2. 2., 2. 3. and 2. 4., respectively). Finally, we use full potential calculations to study the effect of SOC on electronic correlation (Sec. 3.). Sec. 4. is devoted to the summary of results as a conclusion.

1. **Computational Details**

Our calculations are based on two approaches. The first one is using pseudopotential and the second one is a full potential method.

At the first approach, the ab initio non-collinear spin-polarized calculations are done by the Quantum ESPRESSO (QE) package [23] using the Perdew-Zunger (PZ) functional within the local density approximation as well as the Perdew-Burke-Ernzerhof (PBE) functional within the generalized gradient approximation (GGA). To treat the electron-nucleus interaction, the ultrasoft (US) pseudopotentials were employed. We assume an oxidation state of +3 for the Sm element and treat their 4f orbitals as core orbitals. Hence, the atomic valence states are as follows: Sm:5s 5p 5d 6s, Fe: 3s 3p 3d 4s, As: 4s 4p, and O: 2s 2p. Based on experimental and theoretical data, the magnetic phase of SmFeAsO is determined to be the single-stripe-type antiferromagnetic (AFM) [14, 20, 24, 25]. The AFM ordering is presented in Fig. 1, where the magnetic moments of Fe ions are in AFM ordering along a lattice vector whereas ferromagnetic along b lattice vector. The magnetic moments of Sm ions are AFM along c lattice vector [20, 24, 25]. We use 7×7×4 k-point mesh for Brillouin zone sampling of the $\sqrt{2} \times \sqrt{2} \times 1$ supercell which contains four formula units and a marzari-vanderbilt smearing of 0.136 eV for the self-consistent cycles [20]. In order to search the interplay between crystal structure and SOC as well as magnetic ordering, we performed the calculations under four conditions. At first, SmFeAsO compound is assumed to be non-magnetic with and without SOC effect (named as nomag-soc and nomag-nosoc, respectively). Then, it is considered with AFM ordering in presence and absence of SOC effect (named as mag-soc and mag-nosoc, respectively).

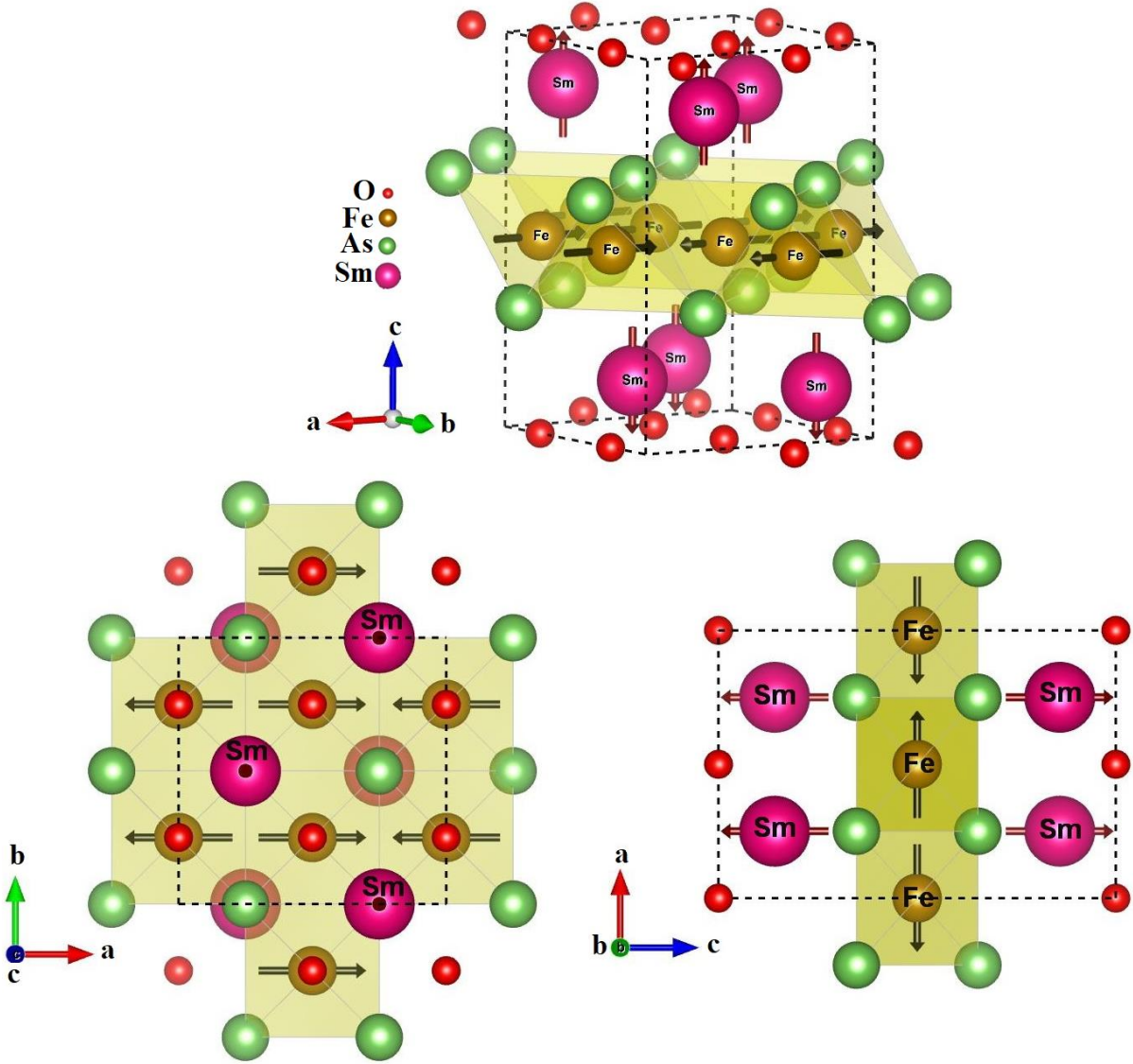

Fig. 1. Crystal and proposed magnetic structure of SmFeAsO based on experiment in the temperature range of $5K \leq T \leq 110K$, shown for a $\sqrt{2} \times \sqrt{2} \times 1$ supercell.

Lattice parameters are determined after full structural optimization through relaxation of both lattice parameters and ionic positions by Broyden-Fletcher-Goldfarb-Shanno (BFGS) method. The geometry relaxations are performed until the total force on each atom becomes smaller than $10^{-3}$ Ry/Bohr in the presence and absence of SOC interaction. The ground state is found to have the ORT crystal structure in agreement with experiment [26]. For studying Tet structure, we force structure to remain Tet during the geometry relaxations.

Since, the optimized bond angles of Fe-As-Fe and As height from Fe layer affect the superconducting transition temperature ($T_C$) [11, 19, 22, 27], we compute them for all above-mentioned calculations. Fig. 2 illustrates the different bond angles of Fe-As-Fe in SmFeAsO compound [19, 27].

We further visualize the calculated Fermi surface in above mentioned different conditions to evaluate the interplay between SOC and the magnetic moments. To illustrate the Fermi surface, the Brillouin zone is sampled by using grids of 14×14×8 k-points mesh based on the Monkhorst-Pack scheme.

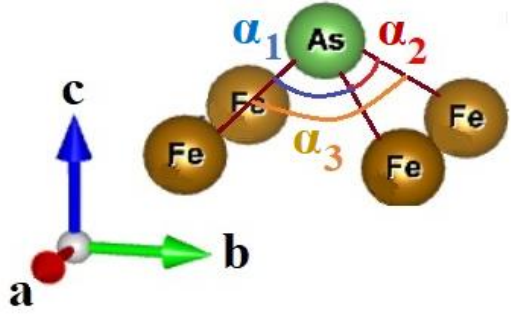

Fig. 2. The different bond angles of Fe-As-Fe.

In the second approach, we employ the state-of-the-art full-potential linearized augmented plane wave (LAPW) method, implemented within elk code [28], to verify the former results from pseudopotential method and survey the effect of SOC on electronic correlation. Since f atomic orbitals of Sm are involved in the calculations, in order to improve Coulomb f-orbital interactions, we use Hubbard U correction (LDA+U) based on around mean field method (AMF) [28, 29]. In this approach, the usual approximate DFT energy functional (LDA) is corrected by an additional term that depends on the effective onsite electron-electron interaction in a similar manner of the U term in the Hubbard model [30]. Moreover, we consider the J parameter in the LDA+U approach which is known as the Hund's rule exchange parameter [2, 28, 31].

Our calculations in this part are based on the optimal constant parameters and ionic positions which are obtained by using a pseudopotential approach in the first step, namely a=3.861, b=3.840 and c=8.108 angstrom. The crystal structure is not relaxed when the U Hubbard and J exchange parameters are applied, as the magnitude of the force acting on each atom remains on the order of $10^{-3}$ Ry/Bohr. For LAPW calculations, we set $k_{max} = 7$ a.u.$^{-1}$, and chose 2.5457, 2.1384, 2.1384 and 1.6365 a.u. for the muffin-tin radius of Sm, Fe, As and O, respectively.

It should be mentioned that the calculations based on GGA approximation with and without SOC are not converged in two approaches, therefore, the calculated results based on LDA approximation are just reported.

## 2. Result and Discussion

### 2. 1. The Crystal Structure

We have performed calculations to determine the optimal lattice parameters of SmFeAsO and the height of the As atoms from the Fe layer under different conditions in both the ORT and Tet crystal structures. The results are summarized in Table. 1 and illustrated in Fig. 3. The LDA functional used in these calculations tends to underestimate the lattice parameters. However, the difference between the a and b lattice constants obtained from the calculations is in good agreement with experimental data [27, 26].

Starting with the experimental data of the Tet structure associated with the superconducting state of SmFeAsO, we conducted a full relaxation with a constraint on the crystal lattice to remain Tet [32]. Our findings indicate that the AFM ordering has a negligible effect on the a and c lattice parameters. Furthermore, we observed that the SOC enhances these parameters to some extent, as shown in Table 1(b). However, when the experimental lattice constants of the ORT structure are used as the starting point for the full relaxation of the crystal structure, we find that the ORT structure becomes the ground state in the presence of AFM ordering. In the absence of AFM ordering, the ORT structure transforms into the Tet structure (Fig. 3 (a)). This transition in the crystal structure occurs through a reduction in the a and c parameters, while the b parameter of the ORT structure increases. Indeed, AFM ordering along a vector enhances a parameter and ferromagnetic ordering along b vector leads to a decrease in the b parameter. The observed modulation and anisotropy in the structure of SmFeAsO when considering AFM order suggest that the tiny orthorhombic ionic distortion is due to magnetoelastic coupling which is the deformation of a material in response to a magnetic field [21]. The comparison of crystal volumes which are included in Table. 1 (a), indicates that the SOC amplifies the magnetostriction. In addition, previous studies have shown that the formation of CDW relies on lattice distortion [33, 34]. Therefore, it can be inferred that AFM ordering, which induces lattice distortion, can give rise to CDW formation [33]. Interestingly, it can be suggested that the CDW in the SmFeAsO compound is derived from a spin density wave (SDW) [35, 36, 37, 38]. In the following sections, additional evidence is presented, including density of states (DOS), electronic band structure, and Fermi surface analysis, to further investigate the coexistence of spin and charge density waves dependently.

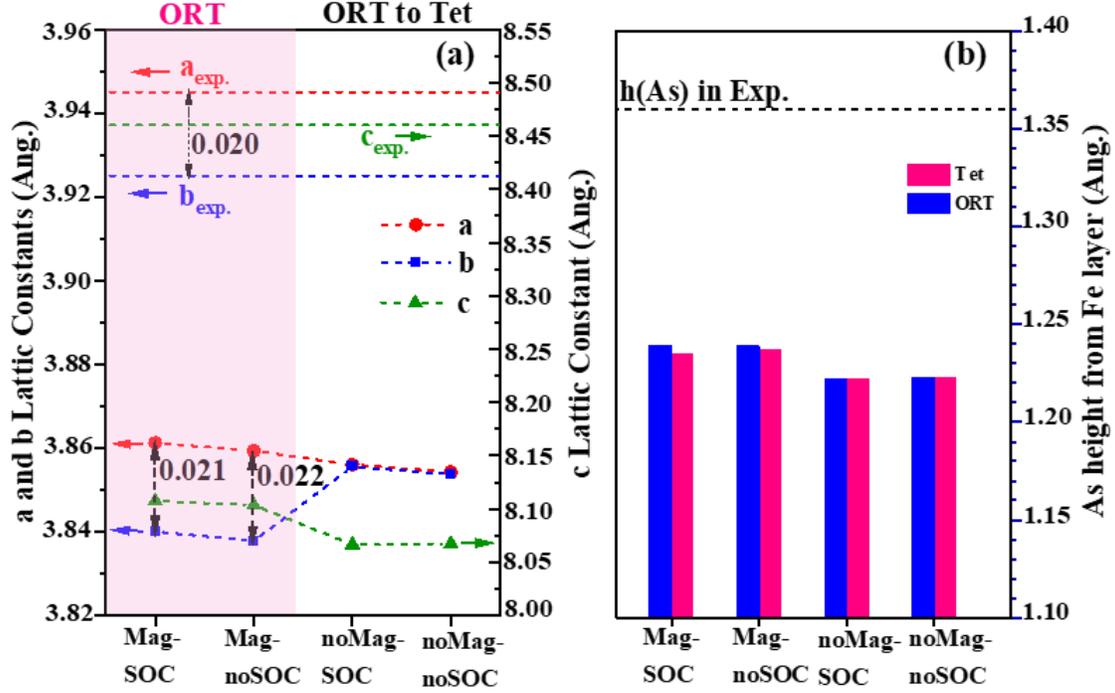

Fig. 3. (a) The Calculated optimal lattice parameters in various conditions are presented for ORT crystal structure which is transformed to Tet by eliminating AFM ordering. On the left side of panel (a), the difference between a and b lattice parameters is indicated by an arrow, and its magnitude is specified. (b) illustrates As height from Fe layer ($h_{As}$) in ORT and Tet. Dashed lines are used to represent the values of related experimental data for comparison (a= 3.945, b= 3.925, c=8.460 and $h_{As}$ = 1.36 Å for ORT) [27, 26].

The calculated As height from the Fe layer ($h_{As}$) in various conditions is depicted in Fig. 3(b). In a detailed investigation, it is observed that AFM ordering increases $h_{As}$. Prior researches have indicated that the critical temperature ($T_C$) of FeAs-based 1111 phase is enhanced by increasing $h_{As}$, with the optimal value achieved for SmFeAsO. This suggests a strong correlation between $h_{As}$ and $T_C$ [19], indicating that AFM ordering can improve the $T_C$ of the superconducting phase. In general, the findings illustrated in Fig. 3 demonstrate that SOC interaction has no noticeable effect on lattice parameters and $h_{As}$ in both crystal structures.

Table. 1 (a) and (b) present the calculated values for different Fe-As-Fe bond angles, magnetic moments on Fe atoms, density of states at the Fermi level, and total energy differences of each phase compared to the nomag-nosoc phase per formula,

Table. 1: The lattice constants, bond angles of Fe-As-Fe, the magnetic moments of different atoms and DOS at Fermi level of SmFeAsO in ORT (a) and Tet crystal structures (b) are calculated through using ultrasoft pseudopotential with LDA functional. The associated experimental data are presented.

| | LDA(US) | | | | Exp. |
|---|---|---|---|---|---|
| | (a) Orthorhombic (ORT) | | | | ORT at T<50K |
| | Mag-SOC | Mag-noSOC | noMag-SOC | noMag-noSOC | Not stable (By doping F atom of 0.01 %, it got stable) |
| a (Å) | 3.861 | 3.859 | 3.856 | 3.854 | 3.945 [26, 32], 3.9408 [20] |
| b (Å) | 3.840 | 3.838 | 3.856 | 3.854 | 3.925 [26, 32], 3.9323 [20] |
| c (Å) | 8.108 | 8.104 | 8.067 | 8.067 | 8.460 [32], 8.4714(2) [20] |
| $V_{unit-cell}$ (Å$^3$) | 120.211 | 120.027 | 119.946 | 119.822 | 130.996 [26, 32], 131.2763 [20] |
| $M_{Fe}$ ($\mu_B$) | 0.550 | 0.560 | 0.000 | 0.000 | 0.34 [32], 0.66(5) [24] |
| $M_{Sm}$ ($\mu_B$) | - | - | - | - | 0.74 [1], 0.62(3) [9] |
| $\alpha_1$ | 72.7 | 72.7 | 73.3 | 73.3 | 70.80 [26] |
| $\alpha_2$ | 73.2 | 73.2 | 73.3 | 73.3 | 71.25 [26] |
| $\alpha_3$ | 114.5 | 114.5 | 115.2 | 115.2 | 110.50 [26] |
| DOS($E_F$) /formula | 1.00 | 0.91 | 2.24 | 2.20 | |
| $E_{Total} - E_{noMag-noSOC}$ /formula (eV) | **-0.682** | -0.005 | -0.677 | 0.000 | |
| | (b) Tetragonal (Tet) | | | | Tet. |
| | Mag-SOC | Mag-noSOC | noMag-SOC | noMag-noSOC | Stable above T=150K |
| a (Å) | 3.856 | 3.853 | 3.856 | 3.854 | 3.9391(2) [32], 3.9427(1) [27] |
| c (Å) | 8.091 | 8.105 | 8.067 | 8.068 | 8.4970(4) [32], 8.4923(3) [27] |
| $V_{unit-cell}$ (Å$^3$) | 120.303 | 120.324 | 119.946 | 119.837 | 131.844(6) [32], 132.012(5) [27] |
| $M_{Fe}$ ($\mu_B$) | 0.516 | 0.537 | 0.000 | 0.000 | - |
| $M_{Sm}$ ($\mu_B$) | - | - | - | - | - |
| $\alpha_1(\alpha_2)$ | 73.1 | 73.0 | 73.3 | 73.3 | - |
| $\alpha_3$ | 114.7 | 114.6 | 115.3 | 115.2 | 110.76(4) [27] |
| DOS($E_F$) /formula | 1.11 | 1.02 | 2.23 | 2.29 | |
| $E_{Total} - E_{noMag-noSOC}$ /formula (eV) | **-0.681** | -0.005 | -0.677 | 0.000 | |

along with the corresponding experimental data for ORT and Tet crystal structures, respectively.

By comparing the computed Fe-As-Fe bond angles of the nomag-nosoc, mag-nosoc, nomag-soc, and mag-soc phases, it is evident that AFM ordering decreases the bond angles, while SOC interaction enhances them in both the ORT and Tet structures. These two factors are in competition, with the effect of AFM ordering being found to be dominant. It was known from experimental studies that the most effective way to increase $T_C$ of Fe-based 1111 superconductors is to decrease the Fe-As-Fe bond angle, thereby reducing the electronic band width [39]. The highest $T_C$ is achieved when the FeAs$_4$ tetrahedron has a perfectly regular structure with a bond angle of 109.47° [19]. Therefore, the AFM ordering can significantly impact the $T_C$ of SmFeAsO, as previously mentioned.

The comparison of the total energy difference of the system under various conditions to the nomag-nosoc phase, reveals that the inclusion of the SOC interaction decreases the total energy and enhances the stability of the system. Although the ground state crystal structure is ORT, there is no noticeable difference in energy between SmFeAsO in the ORT and Tet structures. The energy of the ORT structure is only 1 meV lower than that of the Tet structure.

Furthermore, the calculated magnetic moment on the Fe site decreases by 0.01 $\mu_B$ (0.02$\mu_B$) when the SOC interaction is turned on in ORT crystal structure (Tet), as shown in Table. 1. This suggests that the SOC interaction suppresses the magnetic properties.

It is essential to note that the 4f orbitals of Sm are considered within the pseudo-core, and therefore, the Sm magnetic moment does not contribute to the calculations based on the pseudopotential method.

### 2. 2. Partial and total density of states (PDOS and DOS)

The behavior of the density of states (DOS) at and near the Fermi level is one of the key factors for the occurrence of superconductivity [40]. Therefore, we assess the impact of AFM ordering and SOC on the total DOS and PDOS into the d orbitals and eigenstates of the total angular momentum (j, $m_j$) of Fe d orbitals in both Tet and ORT crystal structures. Since the calculated results of the Tet and ORT structures show no significant differences, we present only the results of the ORT structure under different conditions in Fig. 4.

Fig. 4 (a) displays the DOS and PDOS of the nomag-nosoc phase and shows a considerable peak at 0.3 eV above the Fermi level, which is symmetric to the observed significant peak at -0.3 eV. This could be evidence of the existence of charge density modulation in SmFeAsO [15]. Based Fig. 4 (b) when AFM ordering is considered, the projected DOS on $d_{z2}$ orbital experiences a notable reduction, and the $d_{yz}$ PDOS decreases significantly, accompanied by a partial gap opening at the $E_F$. The $d_{zx}$ and $d_{xy}$ PDOS also diminish partly. Interestingly, according to the DOS analysis, a peak just above the $E_F$ demonstrates significant growth and prominence. This peak shifts towards the $E_F$ level and reaches a value of 0.2 eV. This suggests that the presence of AFM ordering can enhance the charge density modulation.

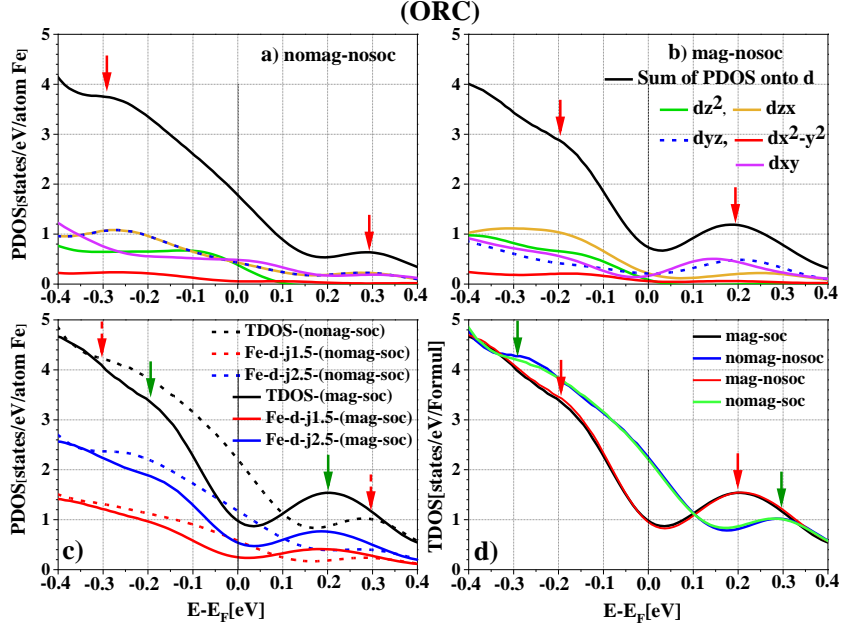

Fig. 4: The projected DOS onto d orbitals of the Fe atom and their sum under different conditions for ORT crystal structure, a) without considering magnetism (nomag-nosoc) and b) with AFM ordering (mag-nosoc) in the absence of SOC effect. c) Total DOS and projected DOS onto the eigenstates of the total angular-momentum $(j,m_j)$ of Fe atom in the present of SOC effect for mag-soc and nomag-soc phases. d) Comparing total DOS of mag-soc, nomag-nosoc, mag-nosoc and nomag-soc phases. The arrows show peaks which can be associated to CDW (SDW).

Fig. 4 (c) illustrates the total DOS and projected DOS onto the eigenstates of the total angular-momentum $(j,m_j)$ of Fe atom in the present of SOC effect for mag-soc and nomag-soc phases. Once again, it is evident that AFM ordering results in electronic localization and the emergence of an outstanding peak near the Fermi level, accompanied by a partial gap opening at the $E_F$. These features indicate the existence of SDW fluctuations, which can induce CDW in the magnetic phase [15, 38, 41, 42]. In Fig. 4 (d) total DOS of mag-soc, nomag-nosoc, mag-nosoc and nomag-soc phases are compared. It reveals that the SOC interaction does not have a considerable effect on the DOS.

The Table. 1 summarizes the DOS at the $E_F$ under different conditions. It is observed that the DOS is decreased with the inclusion of AFM ordering, and further reduced at the $E_F$ with the inclusion of SOC interaction.

The area under the total DOS curve between the Fermi energy and the first maximum peak energy above the Fermi level has been calculated and found to be 0.23 for the mag-soc phase. This value determines the number of electronic states per unit cell and notably is almost equal to the optimal electron-doping concentration (x=0.2) of $SmFeAsO_{1-x}F_x$ achieved by introducing F atoms, which results in the highest transition temperature to the superconducting phase ($T_C$=55 K).

### 2. 3. The Electronic Band Structure

To find out the influence of SOC, AFM ordering and crystal structure on the electronic properties, we have calculated the electronic band structures of SmFeAsO under various conditions. Fig. 5 (a) demonstrates that the SOC interaction reduces degeneracies at boundary edges of Brillouin zone, particularly at $\Gamma$ and $Z$ symmetry points. Based on the orbital-resolved band structures reported by H. O. Choi et al.

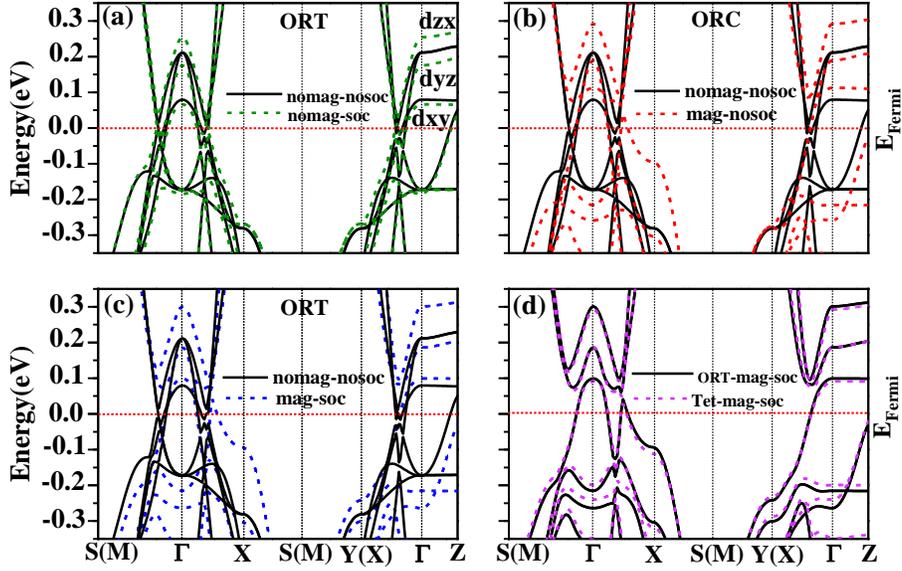

Fig. 5: Comparing between electronic band structures of nomag-nosoc phase and (a) nomag-soc, (b) mag-nosoc, (c) mag-soc phases in ORT crystal structure. The compatibility of band structures of ORT-mag-soc and Tet-mag-soc phases are evaluated in (d). Band structures are plotted along high-symmetry lines in first Brillouin zone of the $\sqrt{2} \times \sqrt{2} \times 1$ supercell. The bands with $d_{zx}$, $d_{yz}$ and $d_{xy}$ orbital characters are determined at $\Gamma$ and $Z$ symmetry points.

for SmFeAsO, and the analysis of the PDOS curves (in the previous section), it is observed that the SOC interaction shifts the $d_{zx}$ band to higher energy and moves the $d_{yz}$ and $d_{xy}$ bands to lower energy. Considering AFM ordering leads to occurring strong anti-crossing and opening orbital-dependent energy gaps in the $d_{yz}$ and $d_{zx}$ bands at the $E_F$ level (Fig. 5 (b)). However, the $d_{xy}$ band remains relatively unchanged and only shifts to higher energy. Fig.5 (c) illustrates the competition between the effects of AFM ordering and SOC interaction on the band structure, simultaneously, where it is evident that AFM ordering dominates over the SOC effect. Fig. 5 (d) further indicates that the transition from the ORT to Tet crystal structure has negligible impact on the electronic band structure.

For a more comprehensive comparison, the energy shifts of the $d_{zx}$, $d_{yz}$ and $d_{xy}$ bands relative to their energies in nomag-nosoc phase are shown in Fig. 6, highlighting the effects of SOC interaction and AFM ordering. These calculations are performed at the Γ and Z symmetry points for both the ORT and Tet crystal structures. It can be observed that the effects of SOC and AFM ordering reinforce each other for the $d_{zx}$ and $d_{yz}$ bands, while they act in an inverse and competitive manner for the $d_{xy}$ band. Therefore, it can be inferred that the behavior of electrons in the $d_{xy}$ state differs from that of electrons in the $d_{zx}$ and $d_{yz}$ bands. However, the effect of AFM ordering dominates over the SOC interaction for the $d_{xy}$ band.

In the subsequent discussion, we will concentrate on the yields of the ORT crystal structure, given the significant similarity between the Tet and ORT results.

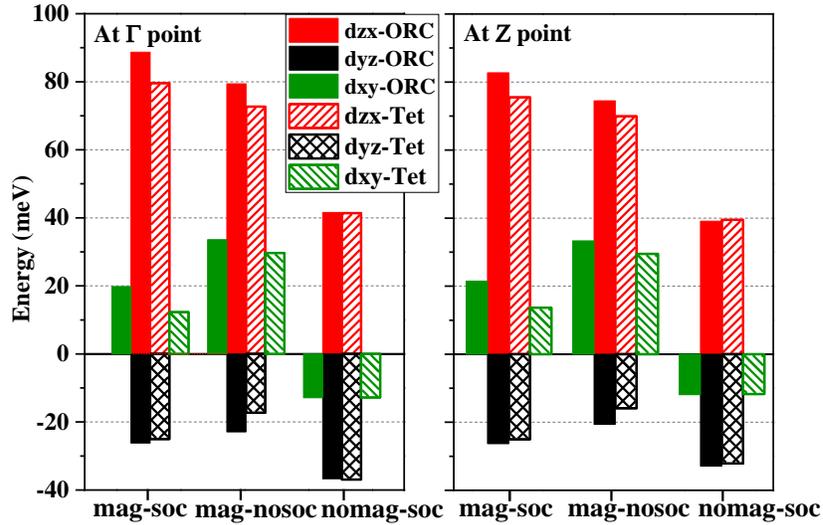

Fig. 6: The energy shifts of $d_{xy}$, $d_{zx}$ and $d_{yz}$ orbital-resolved bands with respect to their energies in nomag-nosoc phase are shown as a result of the effect of SOC interaction and AFM ordering for ORT and Tet crystal structures at Γ (left side) and Z (right side) symmetry points.

Upon analyzing the electronic band structure, DOS and PDOS in the mag-soc phase and comparing it to the nomag-nosoc phase, it is evident that the $d_{yz}$ and $d_{zx}$ bands undergo deformation throughout the entire Brillouin zone, as well as at anti-crossing points. This suggests that the driving mechanism behind the magnetic moments of Fe, with $d_{yz}$ and $d_{zx}$ orbital characters, is not due to Fermi Surface (FS) nesting which is localized in k-space. Conversely, the $d_{yx}$ band remains largely unchanged at the $E_F$ in the presence of AFM ordering and SOC effect. The deformation of the $d_{xy}$ band is primarily confined to the band-crossing points. Based on these findings, it can be concluded that the Fe magnetic moment with $d_{xy}$ orbital character is a result of nesting, and the associated electrons can be considered as itinerant electrons [14, 15]. These results provide confirmation for the hybrid theory of coexisting itinerant and localized electrons. This implies that there are two distinct scenarios: one involving the FS nesting of itinerant electrons, and the other involving the local ordering of a specific degree of freedom [4, 5, 15].

## 2. 4. Fermi Surface

The scanning of the Fermi surface (FS) offers valuable insights into the impact of SOC and AFM ordering on the electronic properties. Fig. 7 visually represents the FS under different conditions. Fig. 7 (a) and (b) display the FS of the nomag-nosoc and nomag-soc phases of SmFeAsO, respectively. The FS of the nonmagnetic phase without SOC effect (Fig. 7 (a)) aligns closely with other theoretical studies [4, 14]. The introduction of SOC interaction leads to the emergence of gaps in the FS, creating deviation in the direction of the Fermi velocity vector from the $k_x k_y$ plane towards the $k_z$ direction at some points as shown in Fig. 7 (b) at $k_z=\pi/c$. When the SOC effect is taken into account, significant changes become apparent, such as the emergence of shallow nodes or near-nodes in the Fermi surface [16]. Fig. 7 (c) and (d) which correspond to the mag-nosoc and mag-soc phases, respectively, demonstrate that in the presence of AFM ordering, most of the bands are shifted away from the FS and the energy gaps are opened. The magnitude of these energy gaps is enhanced by SOC interaction. However, compared to the non-magnetic phase, it has less effect on FS. It is clear that the consideration of SOC interaction and AFM ordering simultaneously leads to a reduction in Fermi surface nesting. Previous studies have shown that the interplay between spin and charge leads to the spontaneous formation of a CDW without relying on repulsive Coulomb interactions or electron-phonon interactions [36]. Therefore, it is possible to predict the coexistence of SDW and CDW in the AFM phase based on specific findings, such as FS nesting, modulation of the electronic density of states and a lattice distortion from Tet to ORT [4, 33, 34, 36].

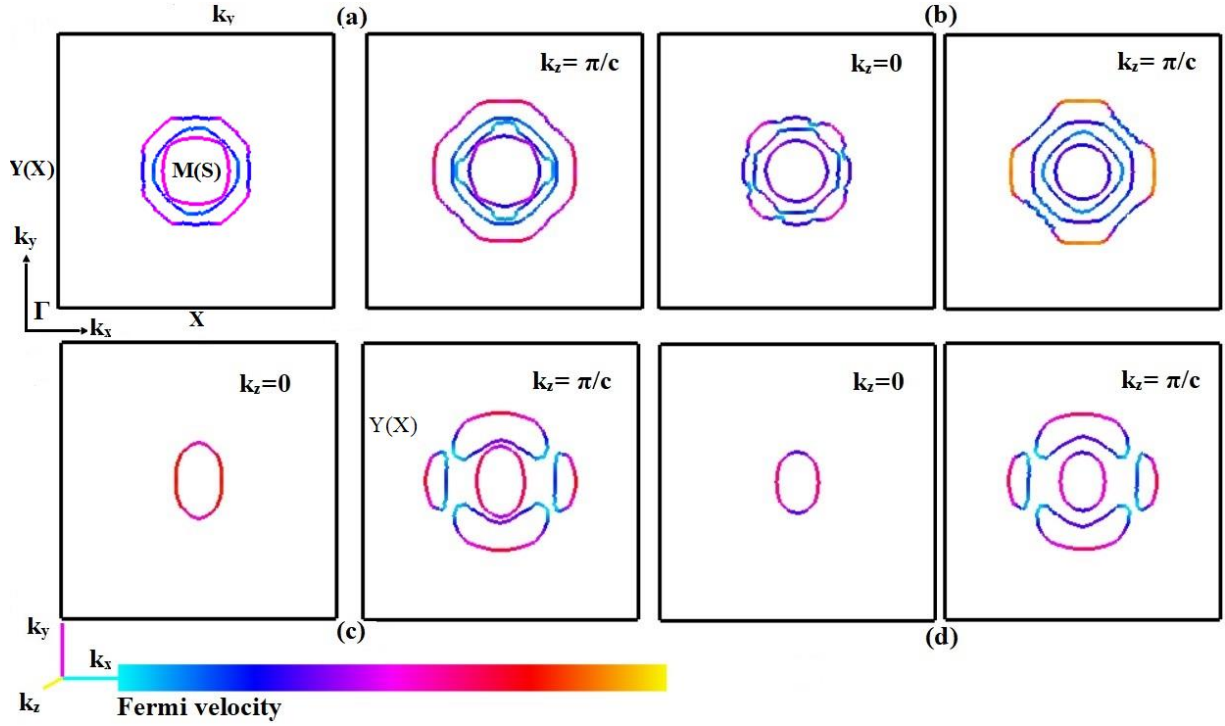

Fig.7 (Color online) Projection of Fermi surface (FS) on $k_z = 0$ and $k_z = \pi/c$ for (a) nomag-nosoc, (b) nomag-soc, (c) mag-nosoc and (d) mag-soc phases of SmFeAsO. Fermi velocity is presented based on the bar color. The FS is illustrated in the first Brillouin zone of the $\sqrt{2} \times \sqrt{2} \times 1$ supercell having four Fe atoms, where $k_x$ and $k_y$ axes are along the x and y axes.

### 3. Results of Using Full Potential Method

In this section, we present the calculated results using the state-of-the-art full-potential linearized augmented plane wave (LAPW) method with the LDA+U+J approach.

Prior to implementing the Hubbard U on Fe and Sm atoms, the magnetic moment calculated using the LDA functional for Fe atoms shows a satisfactory level of agreement with experimental data [32]. However, there is significant deviation between calculated magnetic moment on Sm and the corresponding experimental value (refer to Table. 1). Consequently, we have decided not to apply the Hubbard parameter (U) on the d orbitals of Fe atoms, but it is taken into consideration for the f states of Sm atoms. To determine the appropriate value of U for Sm, we explore a range of U values and assess which one accurately reproduces the experimental magnetic moment observed on the Sm atom [24, 32] and simultaneously removes all f bands from the Fermi energy [34]. In fact, the satisfactory value of U is chosen through a semiempirical approach.

Given the importance of the J parameter in accurately describing the electronic structure of materials with strong spin-orbit coupling, we have reperformed our calculations using J= 0.136 eV (0.01 Ry) as the exchange interaction correction. This allows us to assess the impact of the exchange interaction on the appropriate value of U. Fig. 8 (a) shows the behavior of magnetic moments on Fe and Sm atoms as a function of different U values, considering J=0.00 and 0.136 eV on Sm f-states.

Moreover, the influence of SOC coupling on these calculations is illustrated in Fig. 8 (b). While SOC effect is considered, an anisotropy arises in the magnetic moments of Sm and Fe atoms within AFM ordering [4]. Therefore, we report two absolute values of Sm and Fe magnetic moments, referred to as $Sm_1$, $Sm_2$, $Fe_1$ and $Fe_2$, respectively, where magnetic moment of $Sm_1$ ($Fe_1$) is in the opposite direction to that of $Sm_2$ ($Fe_2$). The magnetic moments calculated for O and As atoms are extremely small, on the order of 0.001 $\mu_B$, so our focus is primarily on magnetic moments of Fe and Sm atoms ($M_{Fe}$ and $M_{Sm}$, respectively).

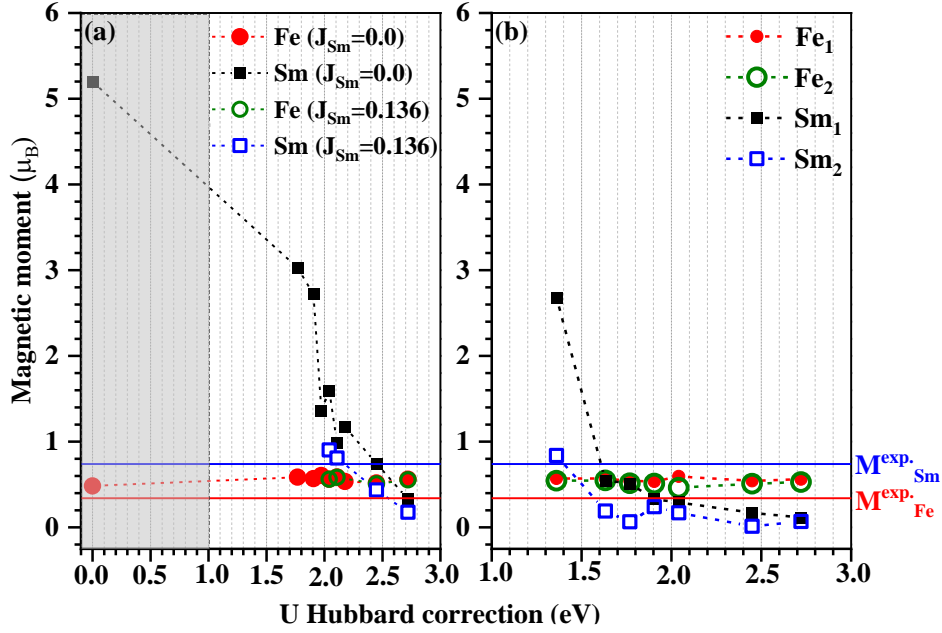

Fig. 8: Magnetic moments on Fe and Sm atoms vs. different Hubbard U correction on f-orbitals of Sm, (a) without SOC interaction for J=0.0 and 0.136 eV as the exchange interaction correction on f-orbitals of Sm, (b) with SOC for J=0.136 eV. The range of Hubbard U in Fig.8 (b) is chosen around the satisfactory value of U obtained in Fig.8 (a) (the range of white window). For better following the manner of magnetic moment on each atom, the calculated points connected to each other by dash line. The reported experimental value of magnetic moments on Sm and Fe atoms are determined through continuous lines [32].

The values of $M_{Fe}$ and $M_{Sm}$ have been previously reported as 0.34 (0.66) $\mu_B$/Fe and 0.74 (0.62) $\mu_B$/Sm, respectively, by Yoichi Kamihara et. al. (Hideo Hosono et. al.) [24, 32]. To achieve agreement with these experimental results, a suitable value of Hubbard U is determined to be 2.449 eV based on Fig.8 (a). When considering the exchange interaction correction, J=0.136 eV, on the f-orbitals of Sm in the calculation, the magnitude of $M_{Sm}$ is reduced, resulting in an appropriate value of U equal to 2.109 eV (as shown in Fig. 8 (a)). It is worth noting that altering the value of J on Sm sites has no impact on $M_{Fe}$.

Figure 8 (b) highlights the effect of SOC on Hubbard U or in other words on electronic correlation. Two obtained magnetic moment values for $M_{Sm}$ and $M_{Fe}$ are illustrated in comparison with experimental data [24, 32].

In this analysis, J is set to 0.136 eV, and the satisfactory value of Hubbard U is determined to be 1.633 eV. It is clear that SOC interaction suppresses the magnitude of $M_{Sm}$ while having no effect on $M_{Fe}$. It suggests that the combination of exchange interaction and SOC reduces electronic correlation, making SmFeAsO a compound with intermediate correlation strength. Since intermediate correlation strength is a fertile ground for superconductivity [43], we anticipate that SOC plays a significant role in the mechanism of superconductivity [44].

The electronic band structure, density of states (DOS), and partial density of states (PDOS) calculated using the satisfactory value of the Hubbard U parameter are in excellent agreement with the results obtained through the pseudopotential approach in the initial analysis.

### 4. Conclusion

In summary, our study focuses on examining the impact of SOC and single-stripe-type AFM ordering on electronic and crystal structure properties of SmFeAsO compound through first-principles study. It is found that the presence of AFM ordering causes a distortion in the crystal structure, transitioning from Tet to ORT. **The Crystal structure transition due to considering AFM order can be attributed to magnetoelastic coupling, which is augmented by considering SOC effect.** Additionally, it is discovered that the introduction of SOC reduces the energy of the ORT structure, making it the preferred ground state. Furthermore, AFM ordering can affect $T_C$ of SmFeAsO in the superconducting phase by increasing As height from Fe layer and reducing the angle of Fe-As-Fe. By evaluating electronic band structure, DOS and PDOS and Fermi surface, we deduce that AFM ordering gives rise to strong anti-crossing and the opening of orbital-dependent energy pseudo-gaps at $E_F$ level and modulation of electronic density of states. SOC interaction has a significant impact on FS nesting, potentially resulting in the creation of shallow nodes or near-nodes on the Fermi surface, particularly in the absence of AFM ordering. However, AFM ordering dominates the SOC effect and induces gap openings at the Fermi surface. Observing Fermi surface nesting, lattice structure distortion, partial gap openings, and the emergence of a noticeable peak near and above the Fermi energy in the electronic density of states provides evidence that the SDW and CDW can coexist interdependently due to AFM ordering. Hence, it can be deduced that the presence of SDW and CDW mutually influences the superconducting mechanism in SmFeAsO. Interestingly, we find that the area under the total DOS curve between the Fermi energy and the first maximum peak above $E_F$ is equal to the optimal electron-doped amount required to induce superconductivity in SmFeAsO. At the end, through full-potential calculations, we determine the satisfactory value of Hubbard U parameter on Sm sites to be 2.449 eV. Considering the exchange interaction correction on Sm atoms (J=0. 0.136 eV) reduces it to 2.109 eV. The magnetic moment on the Sm site is reduced, and anisotropy is induced in the magnetic moments of Fe and Sm atoms at different positions in the ORT crystal structure due to the presence of the SOC interaction. With the inclusion of SOC interaction and exchange correction (J=0.136 eV), the Hubbard U parameter is set at 1.633 eV. It indicates that SmFeAsO exhibits intermediate correlation strength, which is a crucial criterion for superconductivity. Therefore, SOC interaction plays a significant role in the superconducting mechanism of SmFeAsO.

**Acknowledgment**

One of the authors thanks Dr. T. Hashemifar, S. H. Q. Soleimani, E. Agnelli for their invaluable spiritual support.